\documentclass[onecolumn,aps,prl]{revtex4}
\usepackage{graphicx}
[1999/03/29 PSNFSS v.7.2
Times font as default roman
: S Rahtz]

\begin{document}
\renewcommand{\thefootnote}{\fnsymbol{footnote}}
\sloppy
\newcommand{\rp}{\right)}
\newcommand{\lp}{\left(}
\newcommand \be  {\begin{equation}}
\newcommand \ba {\begin{eqnarray}}
\newcommand \bas {\begin{eqnarray*}}
\newcommand \ee  {\end{equation}}
\newcommand \ea {\end{eqnarray}}
\newcommand \eas {\end{eqnarray*}}

\title{Tremor price dynamics in the world's network of stock exchanges}
\thispagestyle{empty}

\author{J\o rgen Vitting Andersen$^1$, Andrzej Nowak$^2$,  
 Giulia Rotundo$^3$ and Lael Parrott$^4$
\vspace{0.5cm}}
\affiliation{$^1$Institut Non Lin\'eaire de Nice
1361 route des Lucioles, Sophia Antipolis 
F06560 Valbonne, France\\ }
\email{vitting@unice.fr}
\affiliation{$^2$Department of Psychology, Warsaw University, 00-183 Warsaw, Poland\\}
\affiliation{$^3$Department of Economics, University of Tuscia, via del 
Paradiso 47, 01100 Viterbo, Italy\\}
\affiliation{$^4$Complex Systems Laboratory, D\'epartment de g\'eographie, 
Universit\'e de Montr\'eal, C.P. 6128 succursale ``centre-ville'' Montreal, Qu\'ebec, Canada}


\date{\today}

\begin{abstract}
{\bf 
We use insight from a model of earth techtonic plate movement to obtain 
a new understanding of the build up and release of stress in 
the price dynamics of the world's stock exchanges. 
Nonlinearity enters the model due to a behavioral attribute of humans
 reacting disproportionately to big changes. This nonlinear response 
allows us to classify price movements of a given stock 
index as either being  
generated due to specific economic news for the country in question, or 
 by the ensemble of the world's stock exchanges 
reacting together like a complex system.  
Similar in structure to the Capital Asset Pricing Model in Finance, 
 the model 
predicts how an individual stock exchange should be priced in terms of 
the performance of the global market of exchanges, but with  
 human behavioral characteristics included in the pricing.  
A number of the model's assumptions are validated against empirical data for 24 of the world's leading stock exchanges.  We show how treshold effects can 
lead to synchronization in the global network of stock exchanges.
 }
\end{abstract}

\maketitle

\vspace{1cm}
\noindent

Like earthquakes,  
financial crises appear to be ever recurrent phenomena with    
the unfolding of a given crisis 
strongly dependent on the history that led up to the crunch.
Whereas the continous build up of stress from techtonic plate 
movements is well understood to be at the origin of earthquakes, 
the causes behind financial distress remain unclear, with 
explanations often sought in singular events. 
Here we present a model that takes a holistic view of how the pricing takes 
place in the world's stock markets. 
 In our model, each stock exchange is represented as a block in a 
network that links any two blocks with a spring of variable strength in 
a world wide network of stock exchanges. As will be explained below, 
the price movements of the stock exchanges are partially created 
by the stick-slip motion of the network of blocks, something very 
similar to ideas 
originally introduced by Burridge and Knopoff (BK)
\cite{Burridge} 
 to describe earthquakes caused by techtonic 
plate movement. This allows a direct study of  
memory effects in the global financial system, with stresses that 
build up over time and are released 
in sudden bursts much like what is seen during seismic activities of 
earthquakes.  
Thus, we emphasize a description 
where the price movements of any given 
stock market can 
not  
be solely understood by looking 
at the level of the individual stock 
exchange and propose that a proper characterisation needs 
to account for system-wide movements at the global level.  

Our objective is to study how "stresses" in the 
global financial system of stock exchanges build up and propagate.  
In our model, stress enters the system because 
of price movements of the indicies represented by 
 displacements 
of the blocks. Stress can either be locally generated due to economic 
news for a specific index, or globally generated due to the transfer of 
stress when a large  
movement happens for a given stock index.  
Similar to the BK model of earthquakes, we assume 
a ``stick-slip'' motion of the indicies so that only a large 
(eventually cumulative) movement 
of a given index has a direct impact in the pricing of the remaining indicies 
world wide. 
In this line of thinking ``earthquakes'' can happen in 
the global financial system because of  
 cascades of big price movements originating from one corner 
of the globe and propagating world wide 
like falling bricks of dominos. We are thus representing the global financial system as a complex system, characterized by important memory effects and path dependence.

A key principle in finance 
 states that as new information is revealed, it immediately 
becomes reflected in the price of an asset and 
thereby loses its relevance\cite{Samuelson,Fama}. 
We suggest to combine this principle 
with a behavioral trait which reflects the tendency of humans
 to reply in a nonlinear fashion to changes, placing 
emphasis on events with big changes and disregarding events with 
modest information content. 
This is in agreement with experiments made in psychology 
which have shown that humans react disproportionally to big changes,  
a phenomenon called change blindness since  
small changes go unnoticed\cite{Jones,Rensink,Lewin}.

As new information is produced at a given exchange, say the opening or 
closing price of that particular market, it becomes part of the 
information that other exchanges may or may not use in their pricing. 
With the existence of futures contracts, this information, as well as 
other economic news, is in principle 
priced in 
 instantaneously, even outside the opening hours of 
exchanges. However if one uses the amount of trading 
volume as a proxy for the relevance of the reaction to new information,  
it is the opening (or respective closing) price that determines 
the most important moment where new information generated prior to the current exchange's trading session becomes priced in. 
Thus, in the following we will use the opening/closing (open/close) prices, which usually correspond to times when the trading volume is highest, as the 
values that become priced in.

Imagine a trader who at the opening of the Tokyo stock exchange tries to 
price in new world wide information coming from the other stock exchanges
about what happened since the markets last closed in Tokyo. We conceive that she/he does
so by taking into account both the release of local economic news in Japan   
(that happened since the previous day's close) 
as well as by seeking out news about how other markets performed 
{\bf after}
 the markets closed in Tokyo. Because of the time zone differences, new 
information at the opening in Tokyo would include the price difference between 
the open and the close the day before 
for the European and American markets.
For the Australian market, however, this would include 
 the price difference between the  
 close of the day before 
and the open the same day, since this 
market is the first market to open world wide, and opens before the 
Japanese markets.  
We postulate a universal behavioral mechanism in the pricing 
 done by traders evaluating 
two different terms  
i) local economic news ii)  
{\em big cumulative} changes from other stock exchanges weighted by their importance 
(in terms of capitalization) and their relatedness (in terms of geographical 
positioning representing e.g., overlap of common economic affairs or  
importance as trading partners). 

At time $t$, the trader of a given stock exchange $i$ estimates the 
price $P_i(t)$ of the index as  
 $P_i(t) = P_i(t-1) \exp{( R_i(t) )}$, with $R_i(t)$ the return 
 of stock exchange $i$ 
between time $t-1$ and $t$:
\ba
\label{R_i}
R_i(t) &  = &  {1 \over N_i^*} \sum_{j \neq i}^{N} \alpha_{ij}
          \Theta (R_j^{cum}(t-1) > R_C) R_j^{cum} (t-1) \beta_{ij} 
           + \eta_i (t),   \\
\label{Rcum}
R_j^{cum}(t) & =  & (1 -
          \Theta (R_j^{cum}(t-1) > R_C) 
[R_j^{cum} (t-1) + R_j(t) ] \\  
\label{N_alpha_beta}
& & N_i^{*}  =    
\sum_{j \neq i}^{N} 
          \Theta (R_j^{cum}(t-1) > R_C) , \ \  
 \alpha_{ij}  =   1 - \exp{\lbrace -K_j/(K_i \gamma ) \rbrace} , \ \  
 \beta_{ij}  =    \exp{\lbrace -(z_i-z_j)/ \tau  \rbrace}  
\ea
$N$ is the total number of stock exchanges. 
The second term in (\ref{R_i}), $\eta_i$, represents internal economic 
news only relevant for 
the specific index $i$, whereas the first term in (\ref{R_i})  
describes external news with large price 
movements of index $j$ having impact on index $i$.  
$t$ stands for the time of the close (respective open) of 
exchange $i$ whereas $t-1$ is the time of the last known 
information (close or open) of exchange $j$ known at time $t$.
 $\alpha_{ij}$ is a 
 coefficient that describes the influence of stock index $j$ on stock 
index $i$ in terms of relative value of the capitalization $K_i,K_j$ 
of the 
two indicies. A large $\gamma (\gamma \gg 1)$ then corresponds to 
 a network of the worlds indicies with dominance of the index with the 
largest capitalization $K_{\rm max}$. Presently this is the U.S. 
financial market, so 
choosing $\gamma$ large corresponds to the case where pricing in any 
country as external information only takes into account the 
movements of the U.S. markets.  
On the contrary a small $\gamma (\gamma \ll 1)$ 
corresponds to a network of indicies with 
equal strengths since $\alpha_{ij}$ becomes independent of $i,j$. 
In addition we assume that countries which are geographically close also 
have larger interdependence economically, as described by the coefficient 
 $\beta_{ij}$ with $z_i-z_j$ the time zone difference of 
countries $i,j$. $\tau$ gives the 
scale over which this interdependence declines. Small $\tau$ ($\tau \ll 1$)
then corresponds to a world where only indicies in the same time zone are 
relevant for the pricing, whereas large $\tau$ ($\tau \ll 1$) describes 
a global influence in the pricing independent of the difference in time zone.  
The structure of (1) is similar to the Capital Asset 
Pricing Model\cite{Treynor} since it  
predicts how an individual stock exchange should be priced in terms of 
the performance of the global market of exchanges, but with  
 human behavioral characteristics included in the pricing.  

$\Theta$ is a Heaviside function so only when the cumulative return of 
index $j$, $R^{cum}_j$, exceeds a threshold return, $R_C$, does the index $j$ 
have a possible (depending on $\alpha_{ij}, \beta_{ij}$) impact on the 
pricing on index $i$. 
The factor $N_i^{*}$ in (1) means that the index $i$ takes into account 
an average impact among the indicies $j$ that have the condition in 
the Heaviside function fullfilled.
(\ref{Rcum}) includes the key 
assumption in Finance that when new 
 information arrives it becomes ``priced in'' in the price 
of an index. That is, after the information that $R_j^{cum}>R_C$ has 
happened, and had an impact on index $i$, this information is deleted ($R_j^{cum} 
\rightarrow 0$). It should be noted however that memory effects are present 
in the model since it is the cumulative ''stress'' that determines when a 
block ''slips''. In Self Organized Critical (SOC) systems, memory is 
known to be an essential ingredient for the criticality of the 
system\cite{Bak}.
Formally  
(1)-(3) describes a 2D BK model of earth 
techtonic plate motion\cite{OFC,Leung}. It can be seen as an 
extension of the 2D Olami-Feder-Christensen (OFC) 
model\cite{OFC} where each block is connected to all 
other blocks 
with $i,j$-dependent coupling constants $C_{ij}=\alpha_{ij}\beta_{ij}$.  
However, in the OFC model each block is only connected to its 4 neighbors and 
has only three ($x,y,z$-dependent) coupling constants. In addition, in our model, ``out of plane'' 
stresses are randomly (in both sign and magnitude) introduced via $\eta_i$ 
at each block instead of the constant (same sign) pull of the OFC model. 
(\ref{R_i}-\ref{Rcum}) gives therefore an interesting perspective of looking at  
the world's financial system as a complex system with self-organizing 
dynamics and possibly similar avalanche dynamics as can be observed 
for earthquakes.

The state of the network of the world's financial markets can according 
to  (1-3) be characterised by the 
4 variables $R_C, \gamma$, 
$\tau$ and the variance 
$\sigma$ of 
$\eta$ 
\cite{note1}. 
 It is possible to use maximum likelihood estimation to 
slave either $\gamma$ or $\tau$ to the remaining 
three parameters\cite{Vitting2}. 
Slaving $\gamma$ to $R_C, \tau$ and $\sigma$ one finds:
\ba 
\label{gamma}
\gamma &  = &  
{
\sum_{t=1}^{T} 
\sum_{i=1}^{N} 
1/N_i^* \lbrack R_i^{data} - \eta_i(t) \rbrack C_i(t-1)
\over
\sum_{t=1}^{T} 
\sum_{i=1}^{N} 
\lbrace 
1/(N_i^{*})^2 C_i(t-1)^2 
- \lbrack \eta_i(t) - R_i^{data}  \rbrack C_i^{'}(t-1) 
 \rbrace 
}  
\\ 
C_i(t) & \equiv  & \sum_{i \neq j}^{N} K_i/K_j \Theta (R_j^{cum}(t-1) > R_C) 
R_j^{cum} (t-1) e^{-(t_i-t_j)/\tau} 
\\ 
C_i^{'}(t) & \equiv  & \sum_{i \neq j}^{N} (K_i/K_j)^2 \Theta (R_j^{cum}(t-1) > R_C) 
R_j^{cum} (t-1) e^{-(t_i-t_j)/\tau} 
\ea 

\begin{figure}[h]
\includegraphics[width=14cm]{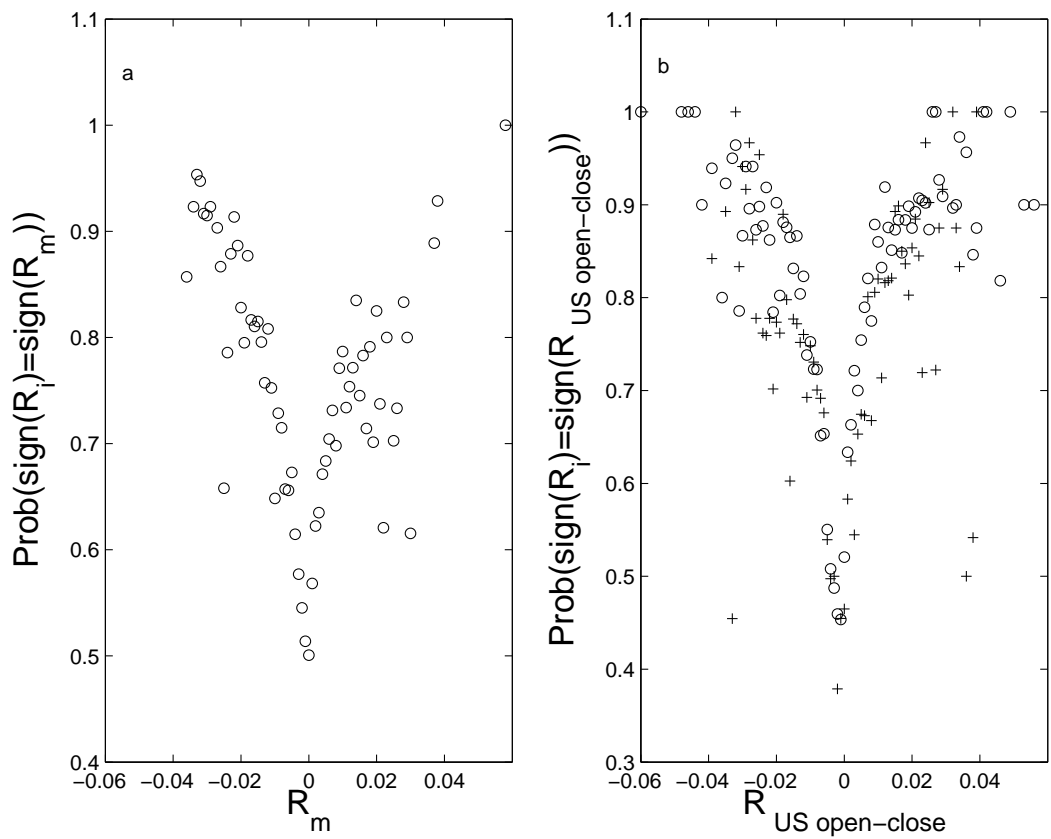}
\caption{\protect\label{Fig1}
a) Conditional probability that the daily return $R_i$ of 
a given country's stock market index has the same sign as the world 
market return defined by $R_m \equiv 
\sum_{j \neq i}^N {K_j R_j \over \sum_{j \neq i} K_j}$ with 
$K_j$ the capitalization of the $j$'th country's index.
b) Conditional probability that the close-open (+: European markets;  
circles: Asian markets) 
return $R_i$  
of  
a given country's stock market index following an U.S. open-close,  
has the same sign as the U.S. open-close return.
}
\end{figure}

To verify the hypothesis that  
 large movements in the stock exchanges play a special role and tend 
to lead to clustering of large movements, we have used empirical data to calculate the 
conditional probability that a given stock market's daily return, $
R=\log{(p(t_{\rm close})/p(t_{\rm close}-1) )}$
has the same sign as 
the daily return of the world market of indicies\cite{data}. 
From Fig.~\ref{Fig1}a it is clear that when the world wide index only exhibits small changes, little coherence is seen between the different 
country's movements.  However,  there appears to be a threshold 
 after which 
large movements in the world wide index lead to sychronization of the individual country exchanges, with the majority tending 
to move in the same direction. Similar results have been found 
 for individual stocks of a given stock market\cite{Cizeau}. This reinforces 
our claim that the stock markets world wide should be considered as {\em one} 
system with {\em large} events playing a special role.

We then checked the specific assumptions in (1-3) that large 
movements of large capital indicies should have a particular 
 impact on smaller 
capital indicies. Using the open-close return of the U.S. stock market gives 
a clear case to check for such a ''large-move'' impact. 
Since the Asian markets close 
before the opening of the U.S. markets, they should only be able to price in  
this information at their opening the following day. An eventual ''large-move'' 
U.S. open-close should therefore have a clear impact on the following close-open 
of the Asian markets. 
On the contrary, the 
European markets are still open when the U.S. market opens up in the morning,  
so the European markets have access to part of the history of the open-close
of the U.S. markets. 
An eventual "large-move'' 
U.S. open-close would therefore still be expected to have an 
impact on the following close-open 
of the European markets, but less so than for the Asian markets since 
part of the U.S. move would already be priced in when the European 
markets closed. Since the opening of the Asian markets by itself could 
 influence the opening of the European markets, this futhermore could 
distort the impact coming from the U.S. markets.   
Figure~1b illustrates again the crucial part of the assumption in our model that 
large moves are indeed special and have impact across markets. 
 As expected, this effect   
 is seen more clearly for the Asian markets compared to the European 
markets.

\begin{figure}[h]
\includegraphics[width=14cm]{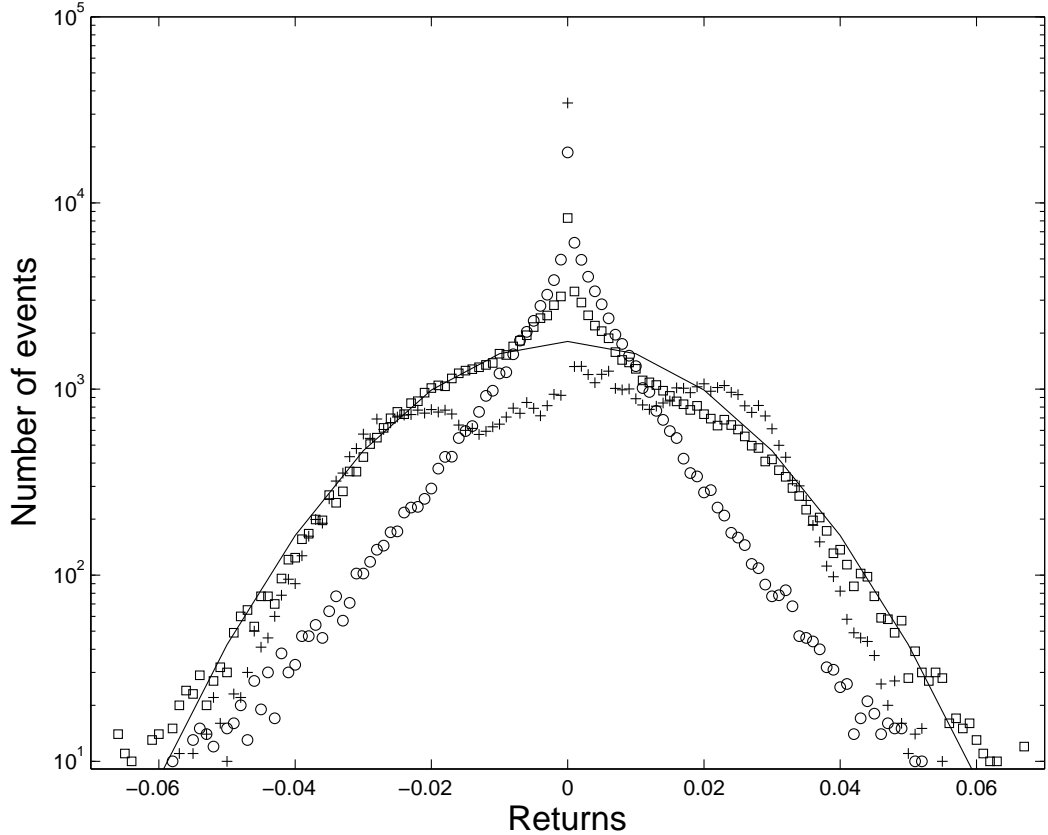}
\caption{\protect\label{Fig2}
Observed returns $R_i$ are shown by circles whereas the 
term $R_i^{\rm transfer} \equiv {1 \over N-1} \sum_{j \neq i}^{N} \alpha_{ij}
          \Theta (R_j^{cum}(t-1) > R_C) R_j^{cum} (t-1) \beta_{ij}$ 
are given by 
$+'s$. Difference  $\eta_i \equiv 
R_i^{\rm transfer} - R_i$ (which according to (1)-(3) 
should be Gaussian distributed) is represented by squares.
Solid line represents  a normal distribution. 
The optimal parameters were found using the central part of the 
distribution of $\eta_i$ taking into account only events which had 
at least 10 occurencies. 
}
\end{figure}

As an additional check on our assumption (1-3)
 we have constructed the difference 
$\eta_i = R_i(t) -  {1 \over N-1} \sum_{j \neq i}^{N} \alpha_{ij}
          \Theta (R_j^{cum}(t-1) > R_C) R_j^{cum} (t-1) \beta_{ij} $ from
the empirical data of 24 of the world's leading stock exchanges using 
daily data since the year 2000. 
According to (1) this difference should be distributed according to a Gaussian 
distribution. We found 
the optimal parameters to be: ($\gamma = 0.8, \tau = 20.0, R_C=0.03, \sigma^2
=  0.0006 $). 
Fig~\ref{Fig2} shows that for these parameter choices,
our definition of price movements 
due to external 
(random) news does indeed fit a normal distribution. 
The obtained values of the optimal parameters suggest a fairly ``global'' 
network of stock exchanges with a large influence of pricing across 
time zones and pricing not only dominated by the largest capital index. 
A priori this seems in agreement with expectations. The value of $R_c$ 
is futhermore consistent with the estimate one can retrieve independently by 
 visual expection of figure~1. 
Lastly, given these optimal parameters, we predicted the sign of 
the open/close for each stock exchange using the sign of 
$R_i^{\rm transfer}$. Using in total 58244 events we 
found a very convincing 63 \% success rate 
of predicting the sign of the return of the open/close of a given stock 
exchange 
{\em ex ante}. 

In analogy with earthquakes, $R_i^{\rm transfer}$ can be thought  
of as describing seismic activity of stress propagating through the system. 
To see if such activity could be used to characterise special 
periods with high "tremor'' activity of the world's stock exchanges 
 we constructed 
$A(t) \equiv \sum_i R_i^{\rm transfer}(t)$. 
As can be seen in figure~3 there is a striking tendency for large 
``tremor'' activity during down periods of the market. That is, 
traders seem to worry more about large movements (positive as well as 
negative) of other countries in the ``bear'' market phase.  

We have introduced a new model of pricing for the world's stock 
exchanges that uses ideas from finance\cite{Vitting} , physics and psychology. 
The model is an extended version of the Burridge-Knopoff  model 
that originally was introduced 
to describe earth techtonic plate movement. 
Like Zumbach et al.\cite{Zumbach} who suggested a 
``Guttenberg-Richter
'' scale of financial market shocks, we have used the analogy with 
earthquakes 
 to get 
a new understanding of the build up and release of stress in 
the world's network of stock exchanges. 
Nonlinearity entered the model due to a
 behavioral attribute of humans 
 reacting disproportionately to big changes. As predicted, 
such a nonlinear response was observed in the impact of 
pricing from one contry to another. 
 The nonlinear response 
allows a classification of price movements of a given stock 
index as either exogeneously  
generated due to specific economic news for the country in question, or 
endogeneously created by the ensemble of the world's stock exchanges 
reacting like a complex system.  
The approach could shed new light of dangers connected to 
systemic risks when large financial ``shocks'' propagate world 
wide\cite{Vitting2}
.

\newpage

\vskip -0.7cm

\begin{figure}[h]
\includegraphics[width=14cm]{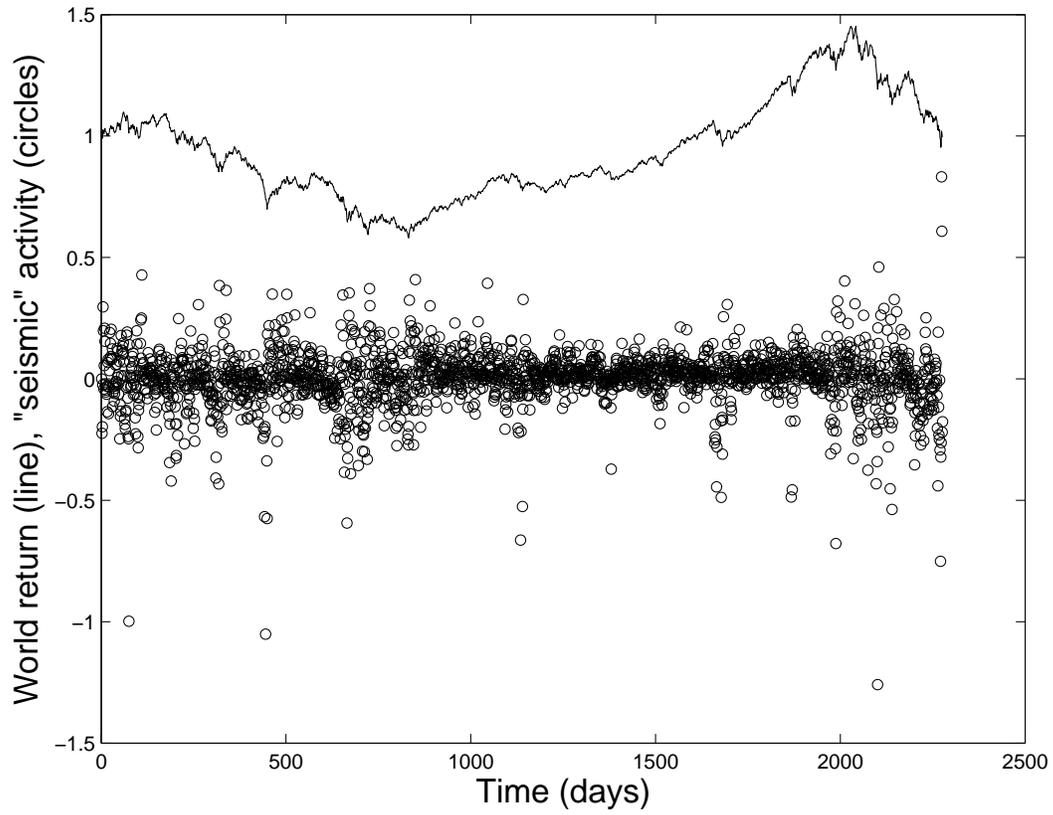}
\caption{\protect\label{Fig3}
Circles represent the term  
$A(t) \equiv \sum_i R_i^{\rm transfer}(t)$ 
 whereas the solid line is the world return index 
normalised according to capitalisation of the different stock indicies. 
}
\end{figure}

{}

\end{document}